\documentclass{sigchi}
\usepackage{balance}
\usepackage{graphics}
\usepackage[T1]{fontenc}
\usepackage{txfonts}
\usepackage{mathptmx}
\usepackage[pdflang={en-US},pdftex]{hyperref}
\usepackage{color}
\usepackage{booktabs}
\usepackage{textcomp}
\usepackage{subfiles}
\usepackage{gensymb}
\usepackage{multirow}
\usepackage{cuted}
\usepackage{capt-of}
\usepackage{microtype}
\usepackage{ccicons}

\def\plaintitle{Accurate and Robust Eye Contact Detection During Everyday Mobile Device Interactions}
\def\plainkeywords{Mobile Phone; Eye Contact Detection; Appearance-Based Gaze Estimation; Attentive User Interfaces}

\makeatletter
\def\@copyrightspace{\relax}
\makeatother

\makeatletter
\def\url@leostyle{%
  \@ifundefined{selectfont}{
    \def\UrlFont{\sf}
  }{
    \def\UrlFont{\small\bf\ttfamily}
  }}
\makeatother
\def\pprw{8.5in}
\def\pprh{11in}

\definecolor{linkColor}{RGB}{6,125,233}
\hypersetup{%
  pdftitle={\plaintitle},
  pdfkeywords={\plainkeywords},
  pdfdisplaydoctitle=true,
  bookmarksnumbered,
  pdfstartview={FitH},
  colorlinks,
  citecolor=black,
  filecolor=black,
  linkcolor=black,
  urlcolor=linkColor,
  breaklinks=true,
  hypertexnames=false
}

\newcommand\Tstrut{\rule{0pt}{2.2ex}}
\newcommand\Bstrut{\rule[-1ex]{0pt}{0pt}}

\begin{document}

\title{\plaintitle}

\numberofauthors{2}
\author{
  \alignauthor{Mihai B\^ace \hspace{.5cm} Sander Staal\\
    \affaddr{Department of Computer Science}\\
    \affaddr{ETH Z\"urich}\\
    \email{\{mbace, staals\}@inf.ethz.ch}}
  \alignauthor{Andreas Bulling\\
    \affaddr{Institute for Visualisation and Interactive Systems, University of Stuttgart}\\
    \email{andreas.bulling@vis.uni-stuttgart.de}}
}

\maketitle

%!TEX root = ../paper.tex

\begin{abstract}
    Quantification of human attention is key to several tasks in mobile human-computer interaction (HCI), such as predicting user interruptibility, estimating noticeability of user interface content, or measuring user engagement.
    Previous works to study mobile attentive behaviour required special-purpose eye tracking equipment or constrained users' mobility. 
    We propose a novel method to sense and analyse visual attention on mobile devices during everyday interactions.
    We demonstrate the capabilities of our method on the sample task of eye contact detection that has recently attracted increasing research interest in mobile HCI.
    Our method builds on a state-of-the-art method for unsupervised eye contact detection and extends it to address challenges specific to mobile interactive scenarios.
    Through evaluation on two current datasets, we demonstrate significant performance improvements for eye contact detection across mobile devices, users, or environmental conditions. 
    Moreover, we discuss how our method enables the calculation of additional attention metrics that, for the first time, enable researchers from different domains to study and quantify attention allocation during mobile interactions in the wild.
\end{abstract}

\category{H.5.m.}{Information Interfaces and Presentation
  (e.g. HCI)}{Miscellaneous} 


%!TEX root = ../paper.tex

\section{Introduction}

With an ever-increasing number of devices competing for it, developing attentive user interfaces that adapt to users' limited visual attention has emerged as a key challenge in human-computer interaction (HCI) \cite{vertegaal2003attentive,bulling2016pervasive}.
This challenge has become particularly important in mobile HCI, i.e. for mobile devices used on the go, in which attention allocation is subject to a variety of external influences and highly fragmented~\cite{Oulasvirta:2005:IBF:1054972.1055101,Steil:2018:FUA:3229434.3229439}.

Consequently, the ability to robustly sense attentive behaviour has emerged as a fundamental requirement for predicting interruptibility (i.e. identifying opportune moments to interrupt a user)~\cite{Choy:2016:LBC:2971648.2971649, Exler:2016:PII:2968219.2968554, Pielot:2017:BIP:3139486.3130956}, estimating the noticeability of user interface content such as notifications~\cite{Pielot:2014:DYS:2556288.2556973}, and for measuring fatigue, boredom~\cite{Pielot:2015:ASD:2750858.2804252}, or user engagement~\cite{McCay-Peet:2012:SAF:2207676.2207751}.

\begin{figure}
	\centering
	\includegraphics[width=\columnwidth]{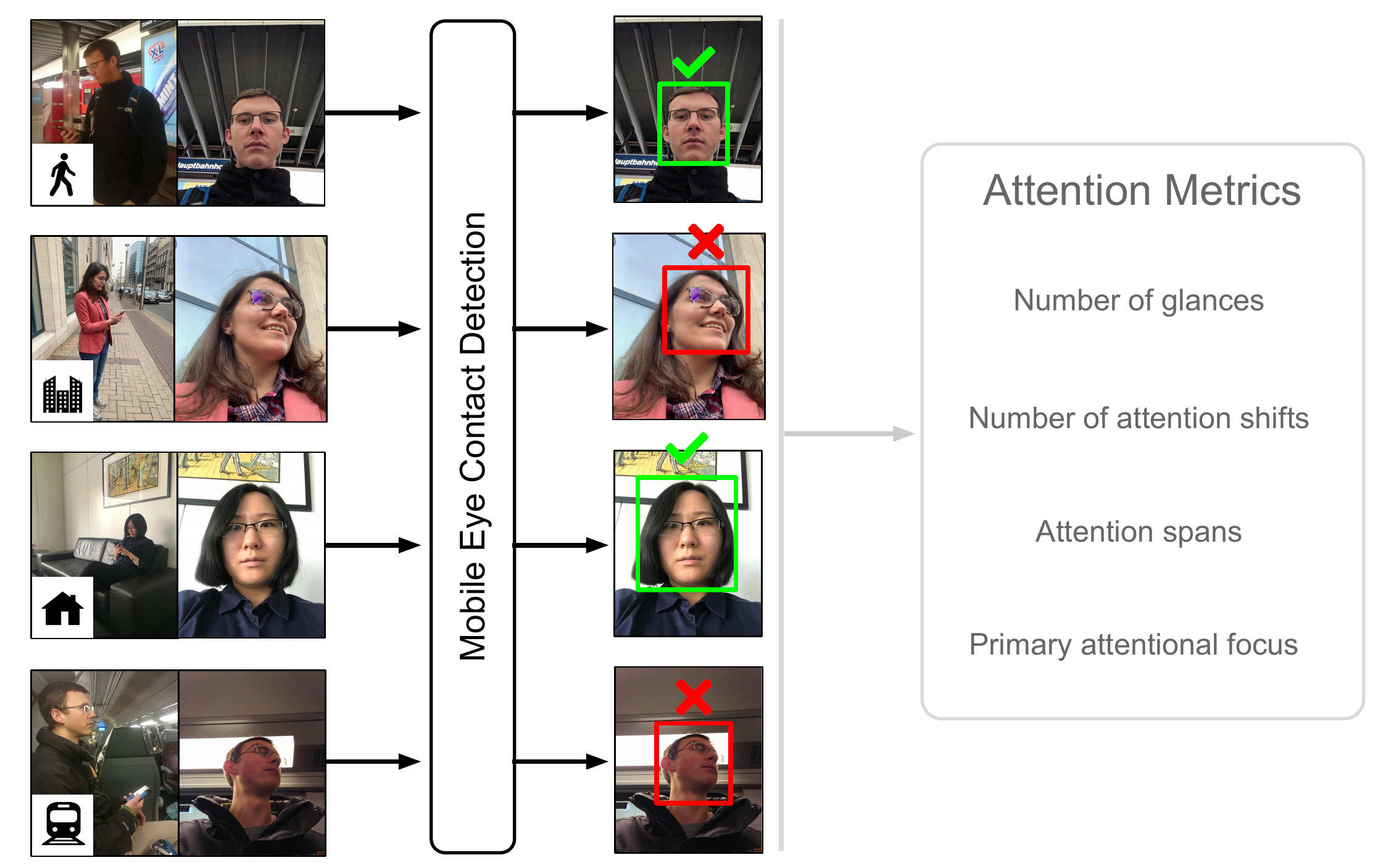}
	\caption{We present a method to quantify users' attentive behaviour during everyday interactions with mobile devices using their integrated front-facing cameras. We evaluate our method on the sample task of eye contact detection and discuss several advanced attention metrics enabled by our method, such as the number of glances, the number of attention shifts, or the average duration of attention span.}
	~\label{fig:teaser}
\end{figure}

Previous methods to sense mobile user attention either required special-purpose eye tracking equipment~\cite{Steil:2018:FUA:3229434.3229439} or limited users' mobility~\cite{10.1007/978-3-319-41627-4_27, DeMasi:2018:YUT:3267305.3267544}, thereby limiting the ecological validity of the obtained findings.
The need to study attention during everyday interactions has triggered research on using device interactions or events as a proxy to user attention, i.e. assuming that attention is on the device whenever the screen is on (e.g. Apple Screen Time) or whenever touch events~\cite{weill2016you,mueller19_etra},  notifications~\cite{Pielot:2014:ISM:2628363.2628364}, or messages~\cite{Dingler:2015:IYQ:2785830.2785840} occur.
While proxy methods facilitate daily-life recordings, it is impossible to know whether users actually looked at their device, and resulting attention metrics are therefore unreliable.
One solution to this problem is manual annotation of attentive behaviour using video recordings~\cite{Oulasvirta:2005:IBF:1054972.1055101} but this approach is tedious, time-consuming, and impractical for large-scale studies.
  
In this work we instead study mobile attention sensing using off-the-shelf cameras and appearance-based gaze estimation based on machine learning.
This approach has a number of advantages.
First, it does not require special-purpose eye tracking equipment given that front-facing cameras are readily integrated into an ever-increasing number of mobile devices and offer increasingly high resolution images.
Second, our approach enables recordings of attention allocation in-situ, i.e. during interactions that users naturally perform in their daily life.
Third, in combination with recent advances in machine learning methods for appearance-based gaze estimation~\cite{zhang19_pami,zhang19_chi} and device-specific model adaptation~\cite{zhang18_chi}, our approach not only promises a new generation of mobile gaze-based interfaces~\cite{khamis18_mobilehci} but also a direct (no proxy required) and fully automatic (no manual annotation required) means to sense user attention during mobile interactions.

We extend a recent method for unsupervised eye contact detection in stationary settings~\cite{zhang2017everyday} and address challenges specific to mobile interaction scenarios:
We use a multi-task CNN for robust face detection~\cite{zhang_face} even on partially visible faces, which is a key challenge in mobile settings~\cite{khamis18_mobilehci}. 
We further combine a state-of-the-art hourglass neural network architecture~\cite{Deng2018CascadeMH} with a Kalman filter for more accurate facial landmark detection and head pose estimation.
Reliable head pose estimates are particularly critical in mobile settings given the large variability in head poses.
We finally normalize the images~\cite{zhang_normalization} and train an appearance-based gaze estimator on the large-scale GazeCapture dataset~\cite{cvpr2016_gazecapture}.

The specific contributions of this work are threefold.
First, we present the first method to quantify human attention allocation during everyday mobile phone interactions.
Our method addresses key challenges specific to mobile settings.
Second, we evaluate our method on the sample use case of eye contact detection and show that our method significantly outperforms the state of the art and is robust to the significant variability caused by mobile settings with respect to users, mobile devices, and daily-life situations on two publicly available datasets~\cite{Khamis:2018:UFE:3173574.3173854,Fathy2015FacebasedAA}. 
Third, we present a set of attention metrics enabled by our method and discuss how our method can be used as a general tool to study and quantify attention allocation on mobile devices in-situ.
%!TEX root = ../paper.tex

\section{Related Work}
\label{sec:relatedwork}
Our work is related to previous works on (1) user behavior modeling on mobile devices, (2) attention analysis in mobile settings, and (3) eye contact detection.

\subsection{User Behavior Modeling on Mobile Devices}
Over the years, smartphones have become more powerful, feature-rich, miniaturised computers. 
Having such devices with us all the time has implications and our usage patterns have changed significantly. 
A study shows that the nature of attentional resources on mobile devices has become highly fragmented and can last for as little as 4 seconds~\cite{Oulasvirta:2005:IBF:1054972.1055101}.
This conclusion is similar to what Karlson et al. identified by looking at task disruption and the barriers faced when performing tasks on their mobile device ~\cite{Karlson:2010:MTC:1753326.1753631}.
Smartphone overuse can have negative consequences in young adults and could lead to sleep deprivation and attention deficits~\cite{Lee:2014:HSE:2556288.2557366}. 
With such changes in the interaction patterns, it has become highly relevant to study and model user behaviour and visual attention.

Sensor-rich mobile devices enable us to collect data and build models with applications in many different domains. 
A significant area of research is concerned with interruptibility or predicting the opportune moments to deliver messages and notifications. 
Mehrotra et al. investigated people's receptivity to mobile notifications~\cite{Mehrotra:2016:MPM:2858036.2858566}.
A different study measured the effects of interrupting a user while performing a task and then evaluated task performance, emotional state, and social attribution~\cite{Adamczyk:2004:EID:985692.985727}.
While many approaches only look at the immediate past for predicting interruptibility, Choy et al. proposed a method which also looks at a longer history of up to one day~\cite{Choy:2016:LBC:2971648.2971649}. 
In a Wizard of Oz study, Hudson et al. analysed which sensors are useful in predicting interruptibility~\cite{Hudson:2003:PHI:642611.642657}.
The way users interact with a certain device does not only depend on the content or the application used, but could be affected by the environment. 
Smartphone usage and interruptibility can also depend on the user's location or social context~\cite{Exler:2016:PII:2968219.2968554, Do:2011:SUW:2070481.2070550}.

Besides looking at interruptibility, others have used attention to model different behavioural traits.
Pielot et al. tried to predict attentiveness to mobile instant messages ~\cite{Pielot:2014:DYS:2556288.2556973}.
User engagement can be analysed by collecting, for example, EEG data~\cite{Mathur:2016:ECM:2971648.2971760} or by looking at visual saliency and how this affects different engagement metrics~\cite{McCay-Peet:2012:SAF:2207676.2207751}.
Toker et al. investigate engagement metrics in visualisation systems which can adapt to individual user characteristics~\cite{Toker:2013:IUC:2470654.2470696}.
Alertness, another indicator for attention, can be monitored continuously and unobtrusively~\cite{Abdullah:2016:CRU:2971648.2971712}.
Such characteristics can be used to better understand user attention patterns.

\subsection{Attention Analysis in Mobile Settings}
Previous research on user behaviour, human attention, or modelling behavioural traits has not focused on mobile devices. 
Given their popularity, understanding, detecting, modelling, and predicting human attention has developed into a new area of research.
VADS explored the possibility of smartphone-based detection of the user's visual attention~\cite{jiang2016vads}.
Users had to look at the intended object and hold the device so that the object, as well as the user's face, can be simultaneously captured by the front and rear cameras.
In an analysis task, knowing where users direct their attention might be sufficient, however, to fulfill the vision of pervasive attentive user interfaces~\cite{bulling2016pervasive}, a system needs to predict where the user's attention will be. 
Steil et al. proposed an approach to forecast visual attention by leveraging a wearable eye tracker and device integrated sensors~\cite{Steil:2018:FUA:3229434.3229439}.
Another approach is to anticipate the user's gaze with Generative Adversarial Networks (GANs) applied to egocentric videos~\cite{zhang2017deep}.
Attention allocation and modelling user attention goes beyond research and lab studies. 
With iOS version 12, Apple has released a built-in feature called Screen Time which measures the amount of time the screen is on and presents usage statistics. 
A similar app from Google for Android is Digital Wellbeing.
Such applications provide interesting insights into one's own usage, however, they are rather naive and always assume the users' attention when the screen is on. 

\subsection{Eye Contact Detection}
Unlike gaze estimation, which regresses the gaze direction, eye contact detection is a binary classification task, i.e. detecting whether someone is looking at a target object or not. 
The first works in this direction used LEDs attached to the target object to detect whether users were looking at the camera or not~\cite{vertegaal2003attentive, Shell:2003:EAD:765891.765981, Dickie:2004:ECS:985921.985927}.
Selker et al. proposed a glass-mounted device which transmitted the user ID to the gaze target object~\cite{Selker:2001:EGE:634067.634176}.
These methods require dedicated eye contact sensors and cannot be used with unmodified mobile devices. 

Recent works focused on using only off-the-shelf cameras for eye contact detection. 
Smith et al. proposed GazeLocking~\cite{Smith:2013:GLP:2501988.2501994}, a simple supervised appearance-based classification method for sensing eye contact.
Ye et al. used head-mounted wearable cameras and a learning-based approach to detect eye contact~\cite{Ye2015}.
With recent advances in appearance-based gaze estimation~\cite{cvpr2016_gazecapture, zhang2017s, Ranjan2018LightWeightHP, 7299081, zhang19_pami}, Zhang et al. proposed a full-face gaze estimation method~\cite{zhang2017s} and introduced an unsupervised approach to eye contact detection in stationary settings based on it~\cite{zhang2017everyday}. 
In their approach, during training, the gaze samples in the camera plane were clustered to automatically infer eye contact labels. 
Extending this method, Mueller et al.~\cite{mueller18_etra} proposed an eye contact detection method which additionally correlates people's gaze with their speaking behaviour by leveraging the fact that people often tend to look at the person who is speaking. 
All these methods were limited to stationary settings and assumed that the camera always has a clear view of the user. 
Only few previous works focused on gaze estimation and interaction on mobile devices but either their performance and robustness was severely limited \cite{wood2014eyetab,holland2012eye} or were studied in highly controlled and simplified laboratory settings \cite{vaitukaitis2012eye}.
As demonstrated in previous works~\cite{Khamis:2018:UFE:3173574.3173854}, these assumptions no longer hold when using the front-facing camera from mobile devices.
\begin{figure*}
  \centering
  \includegraphics[width=\textwidth]{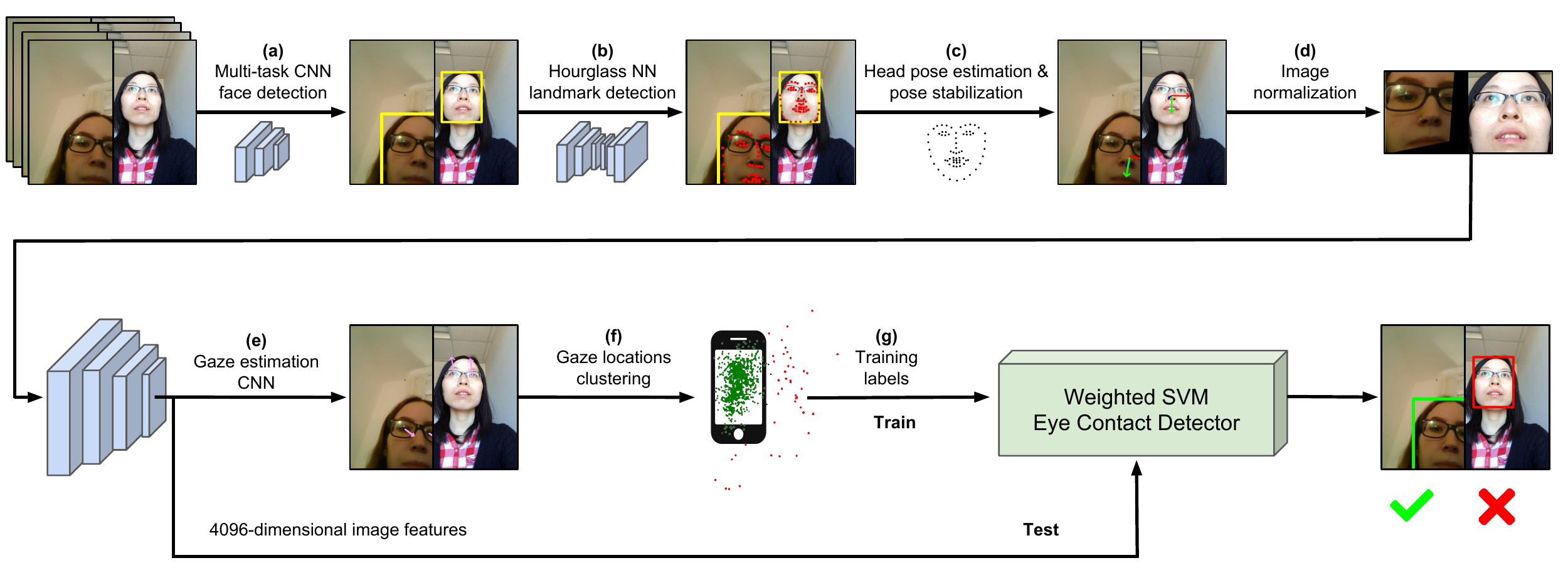}
  \caption{Method overview. Taking images from the front-facing camera of a mobile device, our method first uses a multi-task CNN for face detection~(a) and a state-of-the-art hourglass NN to detect 68 facial landmarks~(b). Then, we estimate the head pose using a 68 3d-point facial landmark model and stabilize it with a Kalman filter (c), a requirement in mobile settings. We then normalize and crop the image (d) and feed it into an appearance-based gaze estimator to infer the gaze direction (e). If the estimated head pose exceeds a certain threshold, we use the head pose instead of the gaze direction. We then cluster the gaze locations in the camera plane (f) and create the training labels (g) for the eye contact detector. The weighted SVM eye contact detector is trained with features extracted from the gaze estimation CNN. 
  }~\label{fig:eye_contact_overview}
\end{figure*}

\section{Method}

To detect whether users are looking at their mobile device or not, we extended the unsupervised eye contact detection method proposed by Zhang et al.~\cite{zhang2017everyday} to address challenges specific to mobile interactive scenarios. 
The main advantage of this method is the ability to automatically detect the gaze target in an unsupervised fashion, which eliminates the need for manual data annotation. 
The only assumption of this approach is that the camera needs to be mounted next to the target object. 
This assumption is still valid in our use case, since the front-facing camera is always next to the device's display.

Figure~\ref{fig:eye_contact_overview} illustrates our method.
During training, the pipeline first detects the face and facial landmarks of the input image. 
Afterwards, the image is normalized by warping it to a normalized space with fixed camera parameters and is fed into an appearance-based gaze estimation CNN. 
The CNN infers the 3D gaze direction and the on-plane gaze location.
By clustering the gaze locations of the different images, the samples belonging to the cluster closest to the origin of the camera coordinate system are labeled with positive eye contact labels and all other samples are labeled with negative non-eye contact labels. 
The labeled samples are then used to train a binary support vector machine (SVM), which uses 4096-dimensional face-feature vectors to predict eye contact.
For inference, the gaze estimation CNN extracts features from the normalized images which is then classified by the trained SVM.

\subsection{Face Detection and Alignment}

Images taken from the front-facing camera of mobile devices in the wild often contain large variation in head pose and only parts of the face or facial landmarks may be visible~\cite{Khamis:2018:UFE:3173574.3173854}.
To address this challenge specific to mobile scenarios, we use a more robust face detection approach which consists of three multi-task deep convolutional networks~\cite{zhang_face}.
In case of multiple faces, we only keep the face with the largest bounding box, since we assume that only one user at a time is using the mobile device. If the detector fails to detect any face, we automatically predict this image to have no eye contact.
After detecting the face bounding box, it is particularly important to accurately locate the facial landmarks since these are used for head pose estimation and image normalization. 
For additional robustness, we use a state-of-the-art hourglass model~\cite{Deng2018CascadeMH} which estimates the 2D position of 68 different facial landmarks.

\subsection{Head Pose Estimation and Data Normalization}

The facial landmarks obtained from the previous step are used to estimate the 3D head pose of the detected face by fitting a generic 3D facial shape model.
In contrast to Zhang et al.\ who used a facial shape model with six 3D points (four from the two eye corners and two from the mouth), we instead used a model with all the 68 3D points~\cite{openface}, which is more robust for extreme head poses, often the case in mobile settings. 
We first estimate an initial solution by fitting the model using the EPnP algorithm~\cite{lepetit2009} and then further refine this solution by doing a Levenberg-Marquardt optimization. 
The final estimation is stabilized with a Kalman filter. 
The PnP problem typically assumes that the camera which captured the image is calibrated. However, since we do not know the calibration parameters of the different front-facing cameras from the mobile devices (nor do not want to enforce this requirement due to the overhead to calibrate every camera), we approximated the intrinsic camera parameters with default values.

Once the 3D head pose is estimated, the face image is warped and cropped as proposed by Zhang et al.~\cite{zhang_normalization} to a normalized space with fixed parameters.
The benefit of this normalization is the ability to handle variations in hardware setups as well as variations due to different shapes and appearance of the face. 
For this, we define the head coordinate system in the same way as proposed by the authors: The head is defined based on a triangle connecting the three midpoints of the eyes and mouth. 
The $x$-axis is defined to be the direction of the line connecting the midpoint of the left eye with the midpoint of the right eye. 
The $y$-axis is defined to be the direction from the eyes to the midpoint of the mouth and lays perpendicular to the $x$-axis within the triangle plane. 
The remaining $z$-axis is perpendicular to the triangle plane and points towards the back of the face.
In our implementation, we chose the focal length of the normalized camera to be 960, the normalized distance to the camera to be 300\,mm and the normalized face image size to be 448 x 488 pixels.

\subsection{Gaze Estimation}

We use a state-of-the-art gaze estimator based on a convolutional neural network (CNN)~\cite{zhang2017s} to estimate the 3D gaze direction. 
Besides the gaze vector, the CNN also outputs a 4096-dimensional feature vector, which comes from the last fully-connected layer of the CNN. 
This face feature vector will later be used as input for the eye contact detector. 
Given that our method was designed for robustness on images captured with mobile devices, we trained our model on the large-scale GazeCapture dataset~\cite{cvpr2016_gazecapture}.
This dataset consists of 1,474 different users and around 2,5 million images captured using smartphone and tablet devices. 
Our trained model achieves a within-dataset angular error of 4.3\degree~and a cross-dataset angular error of 5.3\degree~on the MPIIFaceGaze dataset~\cite{zhang2017s} (which is comparable to current gaze estimation approaches).

To overcome inaccurate or incorrect gaze estimates caused by extreme head poses we propose the following adaptive thresholding mechanism: Whenever the pitch of the estimated head pose is outside the range [$-\theta$, $\theta$], or the yaw outside [$-\phi$, $\phi$], we use the head pose instead of the estimated gaze vector as a proxy for gaze direction.
More specifically, we assume that the gaze direction is the $z$-axis of the head pose.
In practice, we set a value of 40\degree\ for both $\theta$ and $\phi$.

Together with the estimated 3D head pose, the gaze direction can be converted to a 2D gaze location in the camera image plane. 
We assume that each gaze vector in the scene originates from the midpoint of the two eyes.
This midpoint can easily be computed in the camera coordinate system, since the 3D head pose has already been estimated in an earlier step of the pipeline.
Given that the image plane is equivalent to the $xy$-plane of the camera coordinate system, the on-plane gaze location can be calculated by intersecting the gaze direction with the image plane.

\begin{figure*}[h]
	\centering
	\includegraphics[width=\textwidth]{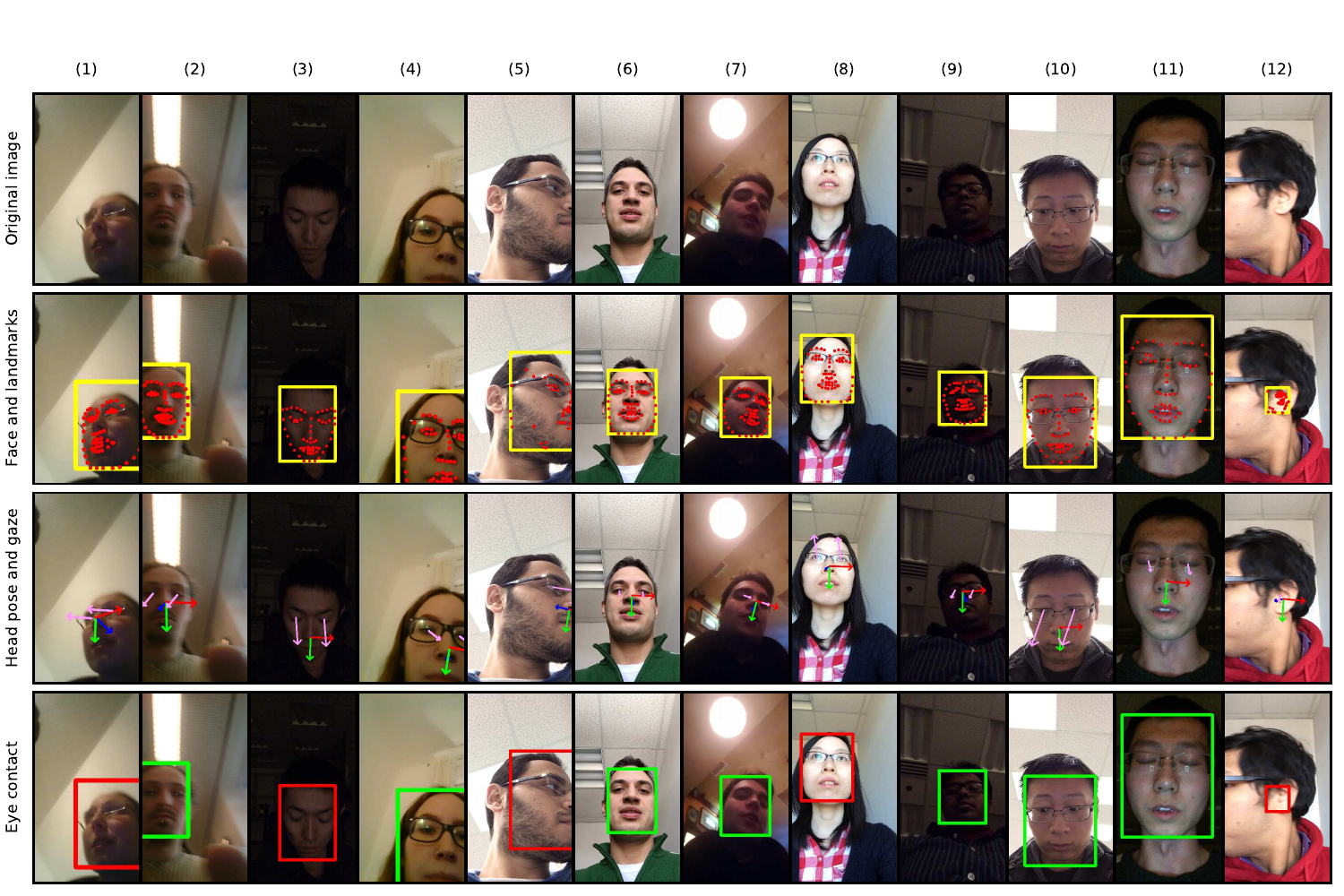}
	\caption{Sample results for eye contact detection on images from the two datasets, MFV and UFEV. The first row shows the input image; the second row the detected face (in yellow) and facial landmarks (in red); the third row shows the estimated head pose and gaze direction (purple); the fourth row shows the eye contact detection result, green for eye contact, red for non eye contact. Columns (1-9) illustrate how our method works across different users, head pose angles, illumination conditions, or when the face is partially visible. Our method can fail if the gaze estimates are inaccurate (10), the eyes are closed (11), or if the face detector fails (12).}
	\label{fig:sample-results}
\end{figure*}

\subsection{Clustering and Eye Contact Detection}

After estimating the on-plane gaze locations for the provided face images, these 2D locations are sampled for clustering.
Similarly, as proposed by Zhang et al.~\cite{zhang2017everyday}, we assume that each cluster corresponds to a different eye contact target object.
In our case, the cluster closest to the camera (i.e., closest to the origin) corresponds to looking at the mobile device. 
To filter out unreliable samples, we skip images for which the confidence value reported by the face detector is below a threshold of 0.9. 
Clustering of the remaining samples is done using the OPTICS algorithm~\cite{ankerst1999}.
As a result of clustering, all the images which belong to the cluster closest to the camera are labeled as positive eye contact samples. 

Finally, taking the labeled samples from the clustering step, we train a weighted SVM based on the feature vector extracted from the gaze estimation CNN. 
To reduce the dimensionality of these high-dimensional feature vectors, we first apply a principal component analysis (PCA) to the entire training data and reduce the dimension so that the new subspace still retains 95\% of the variance of the data. 
At test time, the clustering phase is no longer necessary.
In this case, the 4096-dimensional feature vector is extracted from the appearance-based gaze estimation model and projected into the low-dimensional PCA subspace.
With the trained SVM, we can then classify the resulting feature vector as eye contact or non eye contact.
%!TEX root = ../paper.tex

\section{Evaluation}

We evaluated our method on the sample task of eye contact detection -- a long-standing challenge in attentive user interfaces that has recently received renewed attention in the research community~\cite{zhang2017everyday,mueller18_etra,Ye2015,Dickie:2004:ECS:985921.985927} but so far remains largely unexplored in mobile HCI.
We conducted experiments on two challenging, publicly available datasets with complementary characteristics in terms of users, devices, and environmental conditions (see Figure~\ref{fig:sample-results}): the Mobile Face Video (MFV)~\cite{Fathy2015FacebasedAA} and the Understanding Face and Eye Visibility (UFEV) dataset~\cite{Khamis:2018:UFE:3173574.3173854}. 
We aimed to investigate the performance of our method on both datasets and to compare it to the state-of-the-art method for eye contact detection in stationary human-object and human-human interactions~\cite{zhang2017everyday}.

\subsubsection{Mobile Face Video Dataset (MFV)}
This dataset includes 750 face videos from 50 users captured using the front-facing camera of an iPhone 5s. 
During data collection, users had to perform five different tasks under different lighting conditions (well-lit, dim light, and daylight).
From the five tasks available in the dataset, we selected the ``enrollment'' task where users were asked to turn their heads in four different directions (left, right, up, and down). 
We picked this task because it enabled us to collect both eye contact and non eye contact data. 
From this subset (1 video per task $\times$ 3 sessions $\times$ 50 users), we randomly sampled 4,363 frames that we manually annotated with positive eye contact or negative non eye contact labels. 
58\% of the frames were labeled as positive and 42\% were labeled as negative samples.
This dataset is challenging because it contains large variations between users, head pose angles, and illumination conditions.

\subsubsection{Understanding Face and Eye Visibility Dataset (UFEV)}
This dataset consists of 25,726 in the wild images taken using the front-facing camera of different smartphones of ten participants.
The images were collected during everyday activities in an unobtrusive way using an application running in the background. 
We randomly sampled 5,065 images from this dataset and manually annotated them with eye contact labels. 
We only sampled images where at least parts of the face were visible (which was the case for 14,833 photos). 
Around 17\% of the frames were labeled as negative and 83\% were labeled as positive eye contact samples.
In contrast to the previous dataset, these samples exhibit a class imbalance between positive and negative labels. 
This dataset is challenging because the full face is only visible in about 29\% of the images.

\subsection{Baselines}
There are different ways to detect eye contact, such as GazeLocking~\cite{Smith:2013:GLP:2501988.2501994} which is fully supervised, or methods that infer the coarse gaze direction~\cite{NIPS2015_5848} or leverage head orientation for visual attention estimation~\cite{Voit:2008:DVF:1452392.1452425}.
However, all of these methods are inferior to the state-of-the-art eye contact detector proposed by Zhang et al.~\cite{zhang2017everyday}.
We therefore opted to only compare our method (\textit{Ours}) to two variants of the latter:

\begin{enumerate}
	\item[(1)] \textit{Zhang et al.}~\cite{zhang2017everyday}. Here, we used the dlib\footnote{http://dlib.net/} CNN face detector, the dlib 68 landmark detector, and we trained a full-face  appearance-based gaze estimator on the MPIIFaceGaze dataset~\cite{zhang2017s}. We replicate the original method proposed by the authors.
		
	\item[(2)] \textit{Zhang et al. + FA}. Here, we replace the dlib face and landmark detector. For face detection, we used the more robust approach which leverages three multi-task CNNs~\cite{zhang_face} which can detect partially visible faces, a challenge and a requirement in mobile gaze estimation. Similarly, we replaced the landmark detector with a newer approach which uses a state-of-the-art hourglass model~\cite{Deng2018CascadeMH} to estimate the 2D location of the facial landmarks. The CNN architecture and trained model were the same as in the first baseline. 
\end{enumerate}

In all experiments that follow, we evaluated performance in terms of the Matthews Correlation Coefficient (MCC).
The MCC score is commonly used as a performance measure for binary (two-class) classification problems.
The MCC is more informative than other metrics (such as accuracy) because it takes into account the balance ratios of the four confusion matrix categories (true positives TP, true negatives TN, false positives FP, false negatives FN).
This is particularly important for eye contact detection on mobile devices.
For example, in the UFEV dataset, from the manually annotated images, 83\% of them are positive eye contact and only 17\% represent non eye contact.
A MCC of +1.0 indicates perfect predictions, -1.0 indicates total disagreement between predictions and observations, and 0 is the equivalent of random guessing. 

\subsection{Eye Contact Detection Performance}
\begin{figure}
	\centering
	\includegraphics[width=\columnwidth]{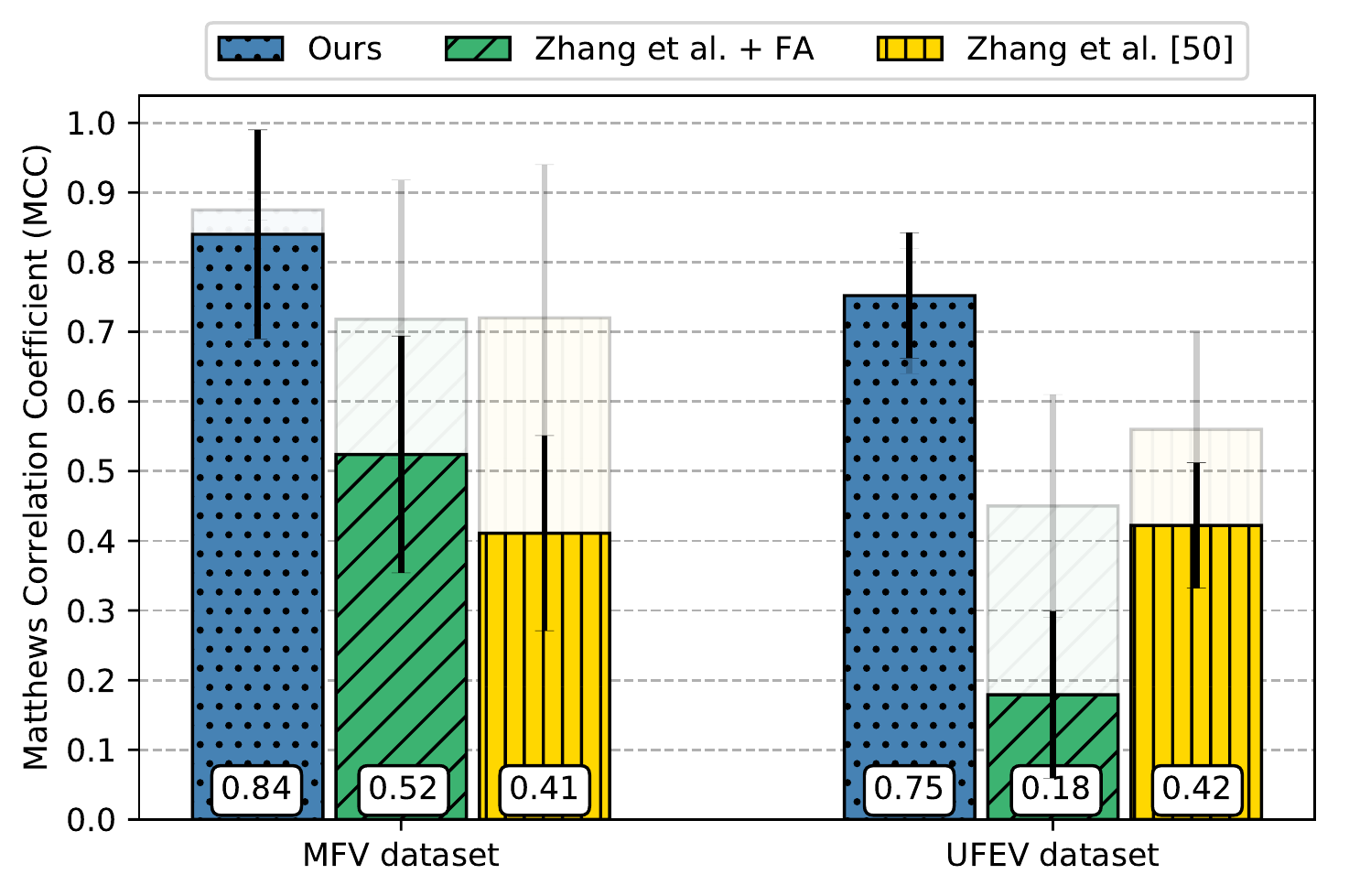}
	\caption{Classification performance of the different methods on the two datasets. The bars are the MCC value and the error bars represent the standard deviation from a leave-one-participant-out cross-validation. The transparent bars illustrate the potential performance improvements when assuming perfect clustering.}
	\label{fig:mcc-overall}
\end{figure}

Figure~\ref{fig:mcc-overall} shows the performance comparison of the three methods.
Our evaluation was conducted on the two datasets using a leave-one-participant-out cross validation. 
The bars represent the mean MCC value and the error bars represent the standard deviation across the different runs. 
As can be seen from the figure, on the MFV dataset our method (MCC 0.84) significantly outperforms both baselines (MCC 0.52 and 0.41).
The same holds for the UFEV dataset where \textit{Ours}~(MCC 0.75) shows significantly increased robustness in comparison to \textit{Zhang et al. + FA}~(0.18) and \textit{Zhang et al.}~(0.42).
The differences between \textit{Ours} and the other baselines are significant (t-test, $p<0.01$).

To better understand the limitations of the clustering and the potential for further improvements, we also analysed the impact of the unsupervised clustering approach on the eye contact classification performance. 
To eliminate the influence of wrong labels resulting from incorrect clustering, we replaced the estimated labels with the manual ground truth annotations.
As such, this defines an upper bound on the classification accuracy given perfect labels.

The transparent bars in Figure~\ref{fig:mcc-overall} show the result of this analysis, i.e. the potential performance increase when using ground truth labels. 
Despite the improvement of the two baselines, our proposed method is still able to outperform them (an MCC score of 0.88 in comparison to 0.72 for both baselines on the MFV dataset and 0.73 in comparison to 0.45 and 0.56 on the UFEV dataset).
Furthermore, our proposed method is close to the upper bound performance with ground truth information. 
We believe this difference can be attributed to our gaze estimation pipeline. 
Due to the improved training steps of the gaze estimator (face and landmark detection, head pose estimation, and data normalization) combined with the GazeCapture~\cite{cvpr2016_gazecapture} dataset, our model can extract more meaningful features from the last fully connected layer of the CNN which, in turn, improves the weighted SVM binary classifier.

\subsection{Performance of Detecting Non-Eye Contact}

The complementary problem to eye contact detection is to identify non eye contact or when the users look away from the device. 
In some datasets, there are only a few non eye contact samples (e.g. only 17\% in the UFEV dataset).
Accurately detecting such events is equally, if not more important and at the same time significantly more challenging than detecting eye contact events due to their sparsity.
One performance indicator in such cases is the true negative rate (TNR).
These events are critical in determining whether there was an attention shift from the device to the environment or the other way around. 
As seen in previous work~\cite{Steil:2018:FUA:3229434.3229439}, these events are not only relevant attention metrics but they can be used as part of approaches to forecast user attention (i.e. predict an attention shift before it actually happens).

\renewcommand{\arraystretch}{1.3}
\begin{table}[t]
	\centering
	\resizebox{\columnwidth}{!} {
    	\begin{tabular}{@{}lccccr@{}}
    
    		\toprule
    		& {\textit{\# images}} & {\textit{GT}} & {\textit{Pred}} & {\textit{TNR}} \Tstrut \\
    		\midrule
    		
    		\multicolumn{5}{@{}l@{}}{\textit{MFV dataset}} \Tstrut\Bstrut \\ \midrule	
    		Zhang et al. & 3,663 & 32.0\% & 7.6\% & 40\% \Tstrut \\
    		Zhang et al. + FA & 3,960 & 36.4\% & 8.8\% & 50\% \Tstrut \\
    		\textbf{Ours} & 3,960 & 36.4\% & 21.5\% & \textbf{88\%} \Tstrut\Bstrut \\ \midrule
            & & & & \\
    		\multicolumn{5}{@{}l@{}}{\textit{UFEV dataset}} \Tstrut\Bstrut \\ \midrule
    		
    		Zhang et al. & 3,517 & 16.4\% & 4.8\% & 51\% \Tstrut \\
    		Zhang et al. + FA & 4,909 & 16.3\% & 9.1\% & 41\% \Tstrut \\
    		\textbf{Ours} & 4,909 & 16.7\% & 8.9\% & \textbf{74\%} \Tstrut\Bstrut \\ \bottomrule
    	\end{tabular}
	}
	\newline
	\caption{Classification performance as true negative rate~(TNR) for non eye contact detection. A comparison between the ground truth negative~(GT) labels distribution and the predicted negative labels distribution~(Pred). The number of images used in the evaluation is dependent on the performance of the face detector. On both datasets, our method is able to outperform the two baselines and can correctly detect more non eye contact events.}~\label{tab:clustering}
\end{table}

Table~\ref{tab:clustering} summarises the results of comparing the TNR of the three methods. 
The TNR measures the proportion of non eye contact (negative) samples correctly identified.
On the MFV dataset, our method is able to outperform the two baselines and identify more than twice as many non eye contact samples~(21.5\% in comparison to 7.6\% or 8.8\%) and more accurately (TNR of 88\%).
On the UFEV dataset the number of predicted samples is comparable for all three methods but our method again significantly outperforms the other two in terms of robustness of identifying non eye contact events~(TNR of 74\% compared to 51\% and 41\% achieved by the other methods). 

\subsection{Cross-Dataset Performance}

In order to realistically assess performance for eye contact detection with a view to practical applications and actual deployments, it is particularly interesting to evaluate the cross-dataset performance.
Cross-dataset performance evaluations have only recently started to being investigated in gaze estimation research~\cite{zhang19_pami} and, to the best  of our knowledge, never for the eye contact detection task.
To this end, we first trained on one dataset, either UFEV or MFV, and then evaluated on the other one. 

\begin{figure}[t]
	\centering
	\includegraphics[width=\columnwidth]{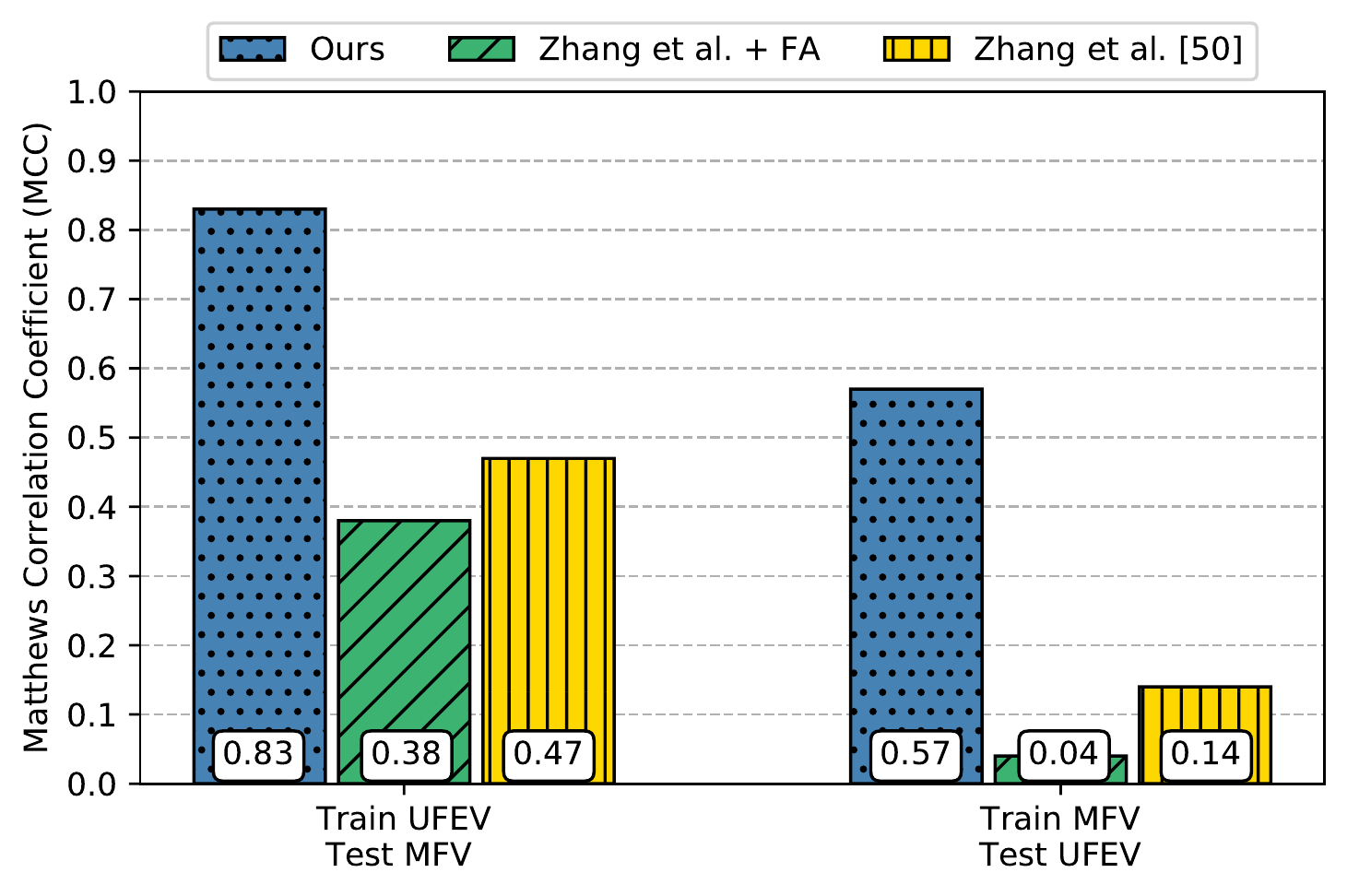}
	\caption{Cross-dataset classification performance of the different methods on the two datasets. In both cases, the three methods were trained on one entire dataset (UFEV or MFV) and tested on the other. The bars represent the MCC value. Our method is able to better abstract away data-specific biases which is important for in-the-wild studies.}
	\label{fig:mcc-cross-dataset}
\end{figure}

Figure~\ref{fig:mcc-cross-dataset} summarises the results of this analysis and shows that our method is able to outperform both baselines by a significant margin both when training on UFEV and testing on MFV, and vice versa.
When training on the UFEV dataset, our method~(MCC 0.83) performs better than the two baselines~(MCC 0.38 and 0.47).
The other way around, training on MFV and testing on UFEV, \textit{Ours}~(0.57) still outperforms \textit{Zhang et al. + FA}~(0.04) and \textit{Zhang et al.}~(0.14).
Taken together, these results demonstrate that our method, which we specifically optimised for mobile interaction scenarios, is able to abstract away dataset specific-biases and to generalize well on other datasets. 
As such, this result is particularly important for HCI practitioners who want to use such a method for real-world experiments on unseen data. 

\subsection{The Influence of Head Pose Thresholding}

\begin{figure}[t]
	\centering
	\includegraphics[width=\columnwidth]{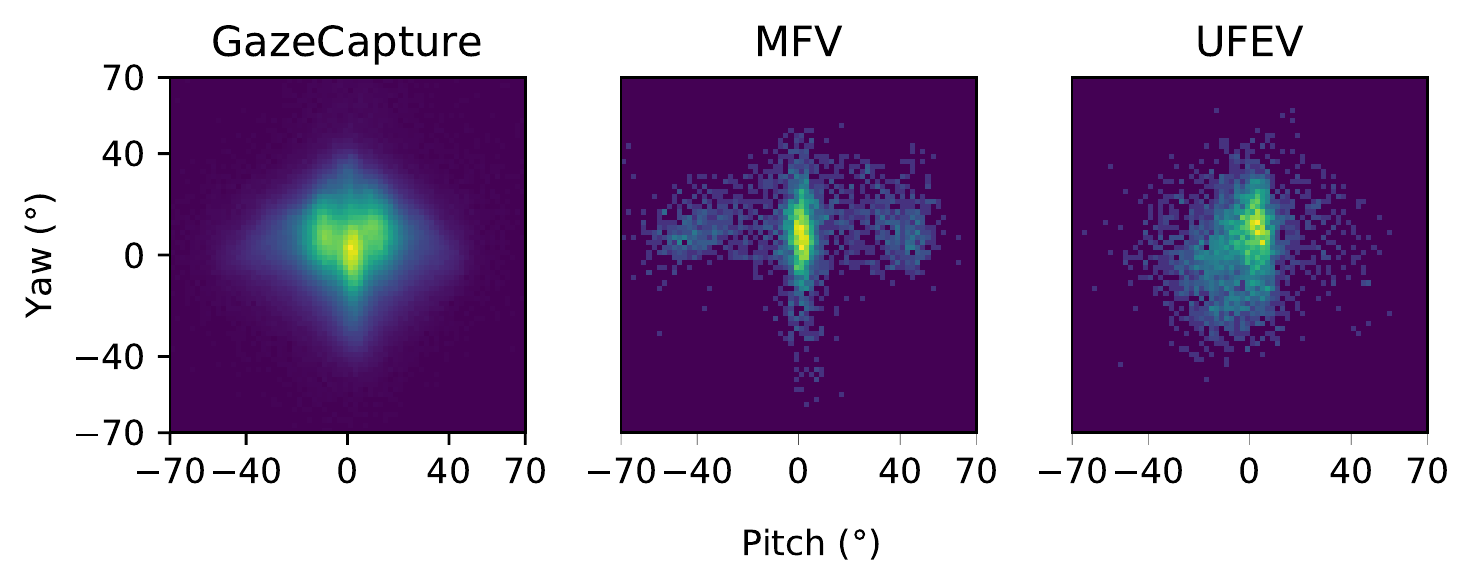}
	\caption{The distribution of the head pose angles (pitch and yaw) in the normalized space for the GazeCapture, MFV, and UFEV datasets. Images collected during everyday activities (MFV and UFEV) exhibit a larger head pose variability than gaze estimation datasets like GazeCapture~\protect\cite{cvpr2016_gazecapture}.}
	\label{fig:norm-headpose-distr}
\end{figure}

In order to reduce the impact of incorrect or inaccurate gaze estimates on eye contact detection performance, in our method we introduced a thresholding step based on the head pose angle. 
Current datasets~\cite{cvpr2016_gazecapture, zhang2017s} have improved the state of the-art in appearance-based gaze estimation significantly, however, they offer limited head pose variability when compared to data collected in the wild (see Figure~\ref{fig:norm-headpose-distr}).
Like in many other areas in computer vision, this fundamentally limits the performance of learning-based methods. 
In our method, we train our model on the GazeCapture dataset which, currently, is the largest publicly available dataset for gaze estimation. 
Still, both the MFV and UFEV dataset show larger head pose variability. 

To overcome the above limitation, we apply the following adaptive thresholding technique: Whenever the horizontal or vertical head pose angle is below or above a certain threshold, we replace the gaze estimates by the head pose angles. 
This adaptive thresholding technique happens in the normalized space~\cite{zhang_normalization}, thus only two threshold values are necessary, one vertical and one horizontal. 

Given the distribution of the GazeCapture training data (see Figure~\ref{fig:norm-headpose-distr}), in our approach we empirically determined a threshold of 40\degree~for the head pose in the normalized camera space.
This is, whenever the head pose angle of either component (vertical or horizontal) is above or below this threshold, we use the head pose angles as a proxy for gaze estimates.

Table~\ref{tab:headpose-thresholding} shows the results of an ablation study with two other versions of our pipeline:
The \textit{Gaze only} (MCC 0.74 on MFV and 0.30 on UFEV) baseline does not use any thresholding.
The \textit{Head pose only} (MCC 0.37 on MFV and 0.73 on UFEV) baseline replaces all the gaze estimates with head pose estimates. 
The results show that \textit{Gaze only} or \textit{Head pose only} can yield reasonable performance for individual datasets.
However, only our method (MCC 0.86 on MFV and 0.76 on UFEV) is able to perform well on both datasets, outperforming both baselines.
This result also shows that, since this value is set in the normalized camera space, the same threshold value is effective across datasets.

\renewcommand{\arraystretch}{1.3}
\begin{table}[t]
	\centering

    	\begin{tabular}{@{}lcc@{}}
    		
    		\toprule
    		 &  \textit{MFV} & \textit{UFEV} \Tstrut \\
     		 \midrule
    		
    		Gaze only & 0.74 & 0.30 \Tstrut\Bstrut \\
    		Head pose only & 0.37 & 0.73 \Tstrut\Bstrut \\
    		\textbf{Ours} (Pitch = Yaw = 40\degree) & \textbf{0.86} & \textbf{0.76} \Tstrut\Bstrut \\ \bottomrule
    	\end{tabular}
	\newline
	\caption{Performance (Matthews Correlation Coefficient) of the three different head pose thresholding techniques on both datasets. \textit{Gaze only} uses no thresholding, \textit{Head pose only} replaces all the gaze estimates by head pose estimates, and \textit{Ours} replaces the gaze estimates by head pose estimates whenever the pitch or the yaw is below or above a threshold.}~\label{tab:headpose-thresholding}
\end{table}

\subsection{Robustness to Variability in Illumination}

\begin{figure}[t]
	\centering
	\includegraphics[width=\columnwidth]{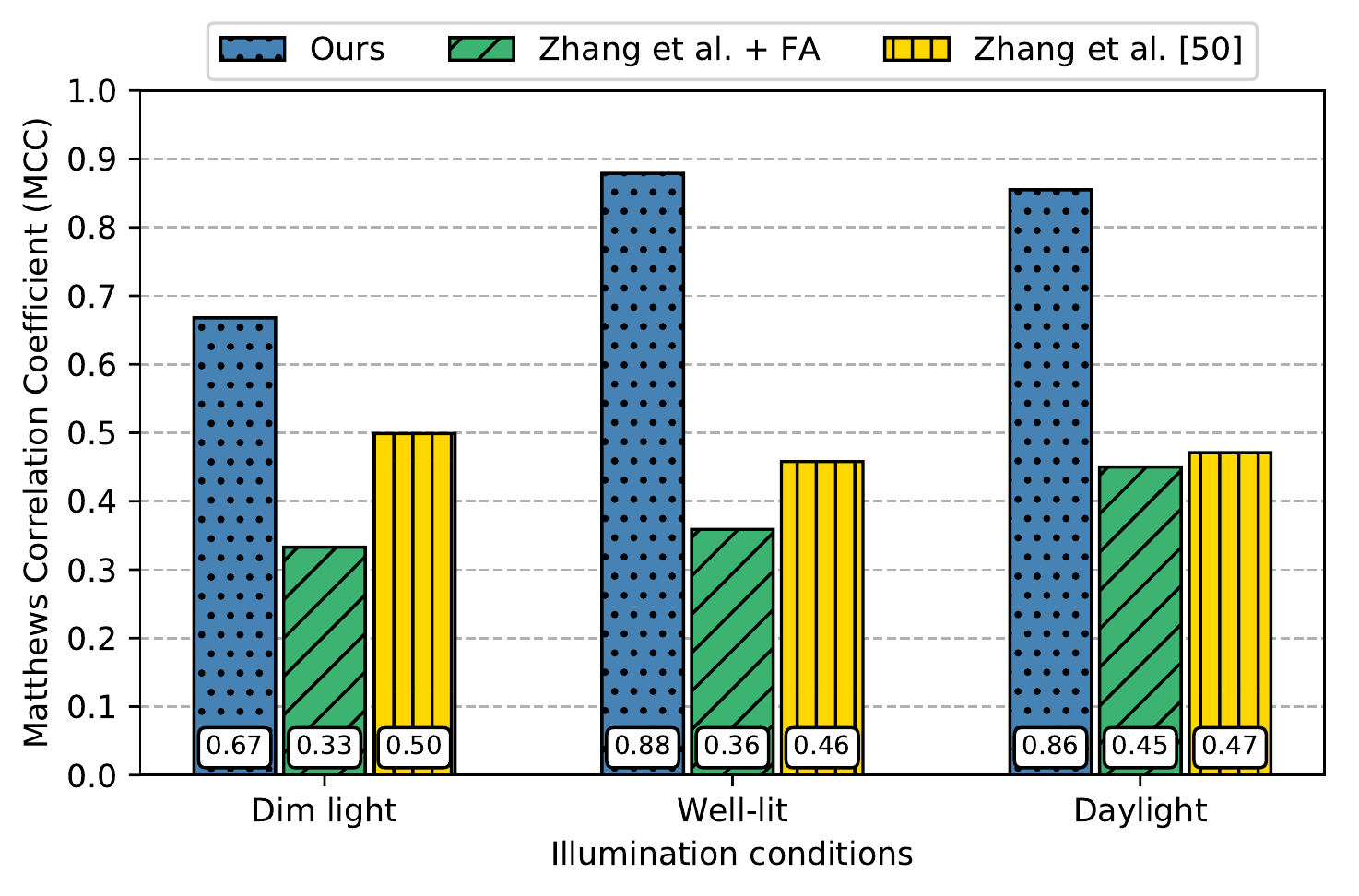}
	\caption{Robustness evaluation of the three methods across different illumination conditions. All three methods were trained on the UFEV dataset and evaluated on the MFV dataset (cross-dataset) in 3 different lighting conditions: dim light, well-lit, and daylight. The bars represent the MCC value. While there is a certain performance drop in dim lighting conditions, our method is consistently more robust and outperforms the two other baselines.}
	\label{fig:mcc-illumination}
\end{figure}

Given that unconstrained mobile eye contact detection implies different environments and conditions, we analysed how varying illumination affected our method's performance in comparison to the two baselines (see Figure~\ref{fig:mcc-illumination}).
In this evaluation, we trained all three methods on the UFEV dataset and evaluated their performance in three different scenarios on a subset from the MFV dataset for which we had both eye contact and illumination labels: \textit{dim light} (986 images), \textit{well-lit} (2157 images), and \textit{daylight} (1221 images).
Our method clearly outperforms the two baselines in all the three scenarios (0.67 vs. 0.50 for dim light, 0.88 vs. 0.46 for well-lit, and 0.86 vs. 0.47 for daylight against the best baseline, \textit{Zhang et. al.~\cite{zhang2017everyday}}).
The \textit{Zhang et al. + FA} baseline is inferior to \textit{Zhang et al.} because it uses the improved face and landmark detector which detects more challenging images otherwise skipped in the evaluation with the \textit{Zhang et al.} baseline.

\section{Discussion}

Our evaluations show that our method not only significantly outperforms the state of the art in terms of mobile eye contact detection performance both within- and cross-dataset (see \autoref{fig:mcc-overall}, \autoref{fig:mcc-cross-dataset}, and \autoref{tab:clustering}) but also in terms of robustness to variability in illumination conditions (see \autoref{fig:mcc-illumination}).
These results, combined with the evaluations on head pose thresholding, also demonstrate the unique challenges of the mobile setting as well as the effectiveness of the proposed improvements to the method by Zhang et al. \cite{zhang2017everyday}.

\begin{figure}[t]
    \centering
    \includegraphics[width=\columnwidth]{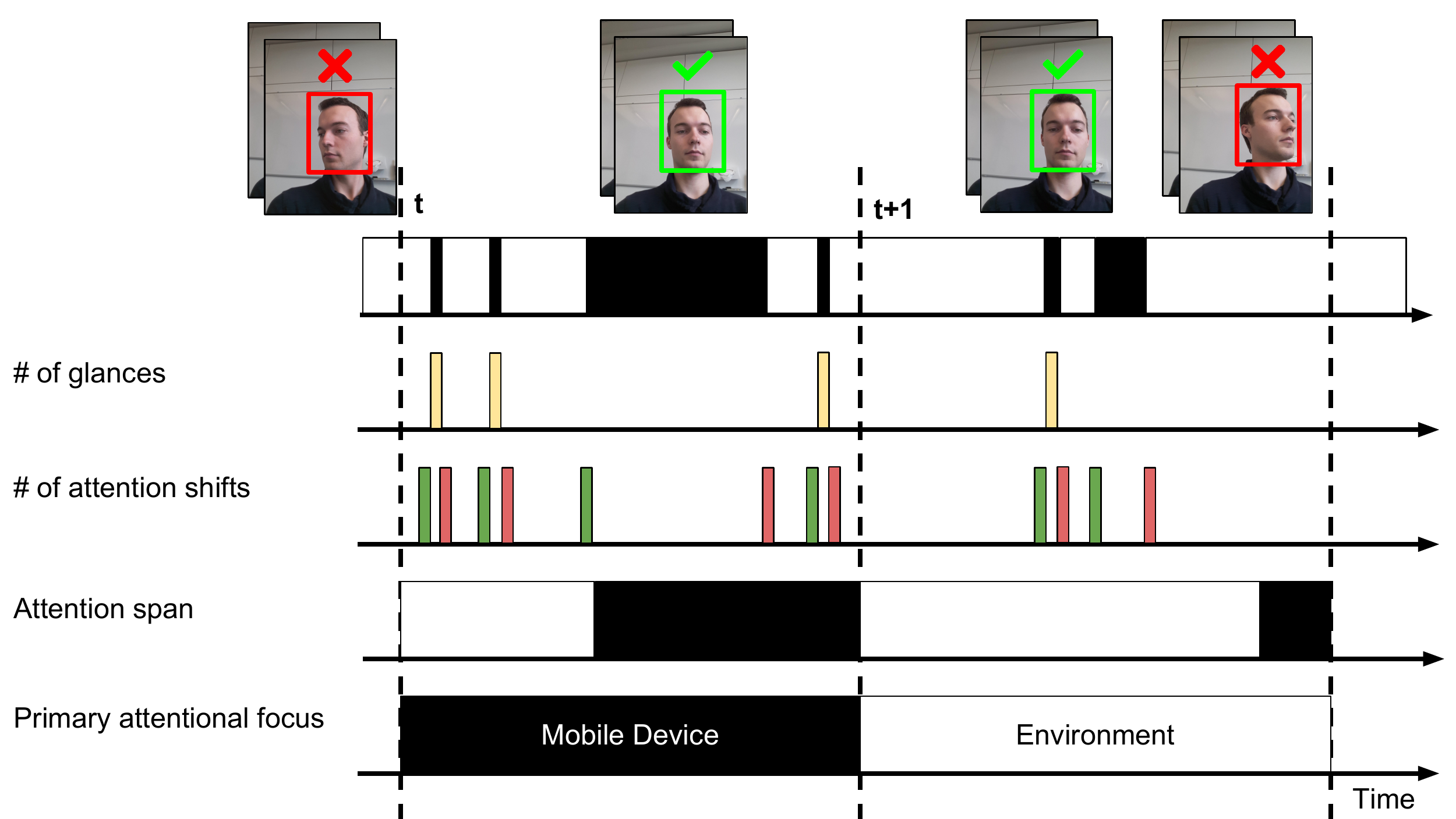}
    \caption{Our eye contact detection method enables studying and quantifying attention allocation during everyday mobile interactions. Knowing when users look at their device (black blocks) and when they look away (white blocks) is a key component in deriving attention metrics such as the number of glances (in yellow), the number of attention shifts (in green from the environment to the device and in purple from the device towards the environment), the duration of attention span (total duration of attention towards the device or the environment in a time interval), or the primary attentional focus.}
    \label{fig:attention-timeline}
\end{figure}

One of the most important applications enabled by our eye contact detection method on mobile devices is attention quantification. 
In contrast to previous works that leveraged device interactions or other events as a proxy to user attention, our method can quantify attention allocation unobtrusively and robustly, only requiring the front-facing cameras readily integrated in an ever-increasing number of devices. 
Being able to accurately and robustly sense when users look at their device, or when they look away, is a key building block for future pervasive attentive user interfaces (see \autoref{fig:attention-timeline}). 

As a first step, in this work we focused on the sample task of eye contact detection.
It is important to note that our method allows to automatically calculate additional mobile attention metrics (see Figure~\ref{fig:attention-timeline}) that pave the way for a number of exciting new applications in mobile HCI.
The first metric that can be calculated is the \textit{number of glances} that indicates how often a user has looked briefly at their mobile device. 
A metric which considers how long users look at their device is the \textit{average attention span}.
In Figure~\ref{fig:attention-timeline}, the average attention span towards the device is given by the average time of the black boxes and the average attention span towards the environment is given by the duration of the white boxes.
Other attention metrics were recently introduced by Steil et al.~\cite{Steil:2018:FUA:3229434.3229439} in the context of attention forecasting.
One such metrics is the \textit{primary attentional focus}:
By aggregating and comparing the duration of all attention spans towards the mobile device as well as the environment we can decide whether the users' attention during the analyzed time interval is primarily towards the device or towards the environment.
Besides aggregating, the shortest or the longest attention span might also reveal insights into users' behaviour. 
Finally, the \textit{number of attention shifts} can capture the users' interaction with the environment.
An attention shift occurs when users shift their attention from the device to the environment or the other way around. 

An analysis which quantifies attentive behaviour with some of the metrics described previously is only the first step. 
Mobile devices are powerful sensing platforms equipped with a wide range of sensors besides the front-facing camera and a user's context might provide additional behavioral insights.
Future work could compare attention allocation relative to the application running in the foreground on the mobile device. 
Such an analysis could reveal, for example, differences (or similarities) in attentive behaviour when messaging, when using social media, or when browsing the internet. 
A different analysis could factor in the user's current activity (attention allocation while taking the train, walking, or standing) or the user's location. 
Going beyond user context, attention allocation could even be conditionally analysed on demographic factors such as age, sex, profession, or ethnicity.

\subsection{Limitations and Future Work}

While we have demonstrated significant improvements in terms of performance and robustness for mobile eye contact detection, our method also has several limitations. 

One of the key components in our pipeline is the appearance-based gaze estimator and our method's performance is directly influenced by it.
In our experiments, we highlighted a limitation of current gaze estimation datasets, namely the limited variability in head pose angles in comparison to data collected in the wild.
As a result, gaze estimates tend to be inaccurate and harm performance of our method.
We addressed this limitation by introducing adaptive thresholding which, for extreme head poses, uses the head pose as a proxy to the unreliable gaze estimates.
Overall, this improved performance but may miss cases when the head is turned away from the device but users still look at it. 
One possibility to address this problem is to collect new gaze estimation datasets with more realistic head pose distributions to improve model training.

Besides further improved performance, runtime improvements will broaden our method's applicability and practical usefulness. 
In its current implementation, our approach is only suited for offline attention analysis, i.e. for processing image data post-hoc. 
While such post-hoc analysis will already be sufficient for many applications, real-time eye contact detection on mobile devices will pave the way for a whole new range of applications completely unthinkable today.
In particular, we see significant potential of real-time eye contact detection for mobile HCI tasks such as predicting user interruptibility, estimating noticeability of user interface content, or measuring user engagement.
Additionally, a real-time algorithm could process the recorded video directly on the device and would not require to store them externally, potentially even in the cloud, as this will likely raise serious privacy concerns.
%!TEX root = ../paper.tex

\section{Conclusion}

In this work, we proposed a novel method to sense and analyse users' visual attention on mobile devices during everyday interactions.
Through in-depth evaluations on two current datasets, we demonstrated significant performance improvements for the sample task of eye contact detection across mobile devices, users, or environmental conditions compared to the state of the art.
We further discussed a number of additional attention metrics that can be extracted using our method and that have wide applicability for a range of applications in attentive user interfaces and beyond.
Taken together, these results are significant in that they, for the first time, enable researchers and practitioners to unobtrusively study and robustly quantify attention allocation during mobile interactions in daily life.

\balance{}

\bibliographystyle{SIGCHI-Reference-Format}
\bibliography{references}

\end{document}